\documentstyle[preprint,floats,prl,aps,epsf]{revtex}

\begin{document}
\draft
\preprint{}

\title{Ray Helicity:  a Geometric Invariant for Multi-dimensional
Resonant Wave Conversion}
\author{Eugene R. Tracy}
\address{Department of Physics, College of William and Mary,\\
Williamsburg, Virginia 23187-8795\\
ertrac@wm.edu}
\author{Allan N. Kaufman}
\address{Lawrence Berkeley National Laboratory and\\ 
Physics Department, UC Berkeley\\
Berkeley, CA 94720\\
ankaufman@lbl.gov}

\date{\today }
\maketitle

\begin{abstract}
For a multicomponent wave field propagating into a multidimensional conversion
region, the rays are shown to be helical, in general.  For a 
ray-based quantity to have a fundamental physical meaning it must be invariant 
under two groups of transformations: congruence transformations (which 
shuffle components of the multi-component wave field) and canonical 
transformations (which act on the ray phase space). It is shown that for 
conversion between two waves there is a new invariant not previously 
discussed:  the {\em intrinsic helicity} of the ray.  

\end{abstract}

\noindent{PACS numbers:  03.65 Sq,42.15 Dp,42.25 Bs}\\

For linear propagation of an $N$-component wave field in a (weakly) non-uniform
multi-dimensional medium, the field is usefully represented by a family of
{\em rays}, along which the wave phase, amplitude, and polarization propagate.
(The theory is referred to as ray-tracing, WKB, or eikonal.)  In a local
region where two waves of different polarization have (nearly) the same
frequency and wave-vector, resonant {\em conversion} occurs, and a ray of one
wave splits into two rays, one of each type.  In a conversion
region, the WKB theory breaks down and a local theory must be developed
which governs the pairwise interaction of the two resonant waves.  Such
a theory is represented by a $2\times 2$ local wave equation.

The ray equations are Hamiltonian with the determinant of the full 
$N\times N$ dispersion matrix playing the role of the Hamiltonian.
In a conversion region, the local ray geometry can be used
to guide the reduction from the original $N\times N$ theory to the local 
$2\times 2$ form~\cite{physletta,tutorial}, construct the local 
transformations which recasts the local $2\times 2$ wave equation into the 
simplest possible form, and find the local coordinates in which the
$2\times 2$ wave equation separates.  The new results
reported here are:  1) In a conversion region, the six-dimensional ray phase 
space is locally foliated by four-dimensional invariant subspaces.  Within 
these four-dimensional subspaces, the ray dynamics is 
hyperbolic in one two-dimensional subspace, and elliptic in the other 
({\em i.e.} pure hyperbolic motion is not possible).  2) There is one 
quantity, not previously identified, that is invariant under all local 
congruence transformations and canonical transformations: the {\em ray 
helicity}.

Linear wave conversion is ubiquitous throughout physics (see, for example,
references given in~\cite{tutorial,robertandgregannals}). While there is a 
large literature on conversion in one dimension, there are
relatively few studies of the multi-dimensional case (see 
Refs.~\cite{bernstein,lazar1,lazar2,lazar3,lazar4,raybasedconversion,robertlandauzener,robertandgregchaos,pre,emission,yuri} in physics, 
and~\cite{braam1,braam2,deverdiere} in mathematics).

We start with a general $N$-component field 
$\Psi =(\psi _1,\psi _2,\ldots ,\psi _N)$ in a weakly non-uniform 
medium. Assume the medium has three spatial dimensions for 
concreteness and is time-stationary, 
but note that the methods described can be extended to include non-stationary 
media as well~\cite{macdonald}.  Denote a point in space as 
${\bf x}=(x_1,x_2,x_3)$.  The wave equation can always be cast into the 
standard form~\cite{lazar2}:
\begin{equation}
\int d^3x'dt'D_{mn}({\bf x},{\bf x}',t-t')\psi _n({\bf x}',t')=0, \; m=1,2\ldots N.
\label{waveequation}
\end{equation}
We assume that the wave kernel $\bf D$ is 
an $N\times N$ matrix-valued fuction of its arguments satisfying 
$D_{mn}({\bf x},{\bf x}',t-t')=D^*_{nm}({\bf x}',{\bf x},t'-t)$, which 
gives non-dissipative wave propagation.   
The goal is to find the multi-component field $\Psi ({\bf x},t)$ throughout a given space-time region with fitting to appropriate initial/boundary 
conditions.  A standard tool for this analysis is the eikonal, or WKB, 
method.  An ansatz is used of the form 
$\Psi ({\bf x},t)=e^{i\theta ({\bf x})-i\omega t}{\tilde \psi }({\bf x})
{\hat {\bf e}}({\bf x})$,
where $\theta ({\bf x})$ is a rapidly varying phase, ${\tilde \psi}({\bf x})$ 
is a slowly varying {\em scalar} amplitude function (assumed real), and 
${\hat {\bf e}}({\bf x})$ is a slowly varying (complex) polarization vector.
The WKB approximation breaks down near caustics and in conversion regions.  
The theory of caustics is well-developed~\cite{robert2,caustics}, while the 
conversion problem in multi-dimensions is our present concern. 

Starting with the wave kernel and using the Weyl 
calculus~\cite{pre,macdonald,Weyl,robert}, we next construct the 
{\em dispersion matrix}, an $N\times N$ matrix-valued function on the ray 
phase space denoted ${\bf D}({\bf x},{\bf k})$.  
From now on we denote a point in the six-dimensional ray phase space by 
${\bf z}\equiv ({\bf x},{\bf k})=(z_1,z_2,z_3,z_4,z_5,z_6)$.
If~(\ref{waveequation}) is conservative (as we assume), then 
${\bf D}({\bf z})$ is Hermitian at each point ${\bf z}$:   
${\bf D}^{\dag }({\bf z})={\bf D}({\bf z})$.  Ray propagation requires  
$D({\bf z})\equiv det({\bf D})=0$.  This single scalar condition upon the 
six coordinates ${\bf z}$ defines the {\em dispersion surface} which is, 
generically, a smooth five-dimensional surface, though it may have local 
singularities.  Conversion occurs in the vicinity of
the spatial point ${\bf x}_*$ where two different WKB waves of frequency
$\omega$, with distinct polarization and dispersion characteristics, have 
nearly equal wavevectors~\cite{tutorial}.   In the ray phase space, 
this requires that two branches of the dispersion surface are in close 
proximity near the point ${\bf z}_*=({\bf x}_*,{\bf k}_*)$.

Away from conversion regions (and caustics), the amplitude and polarization 
vary slowly following a ray.  Within conversion regions, however, the 
polarization and amplitude vary rapidly and the WKB ansatz is no longer 
valid.  Instead, the field locally has the form $\Psi ({\bf x},t)=
e^{i{\bf k}_*\cdot ({\bf x}-{\bf x}_*)-i\omega t}
\left[\psi _{\alpha }({\bf x}){\hat {\bf e}}_{\alpha } +
\psi _{\beta }({\bf x}){\hat {\bf e}}_{\beta }\right]$,
where ${\hat {\bf e}}_{\alpha }$ and ${\hat {\bf e}}_{\beta }$ are the 
{\em uncoupled} polarizations evaluated at the conversion point.  The 
algorithm for finding these two constant polarization vectors has
been discussed elsewhere 
(see~\cite{lazar2,lazar4} for a proposed algorithm).  
We assume the uncoupled polarizations are given. The {\em complex} scalar 
amplitude functions, $\psi _{\alpha }$ and $\psi _{\beta }$, include all 
the effects of the coupling and have rapid variation in both amplitude and 
phase in the conversion region, but (after multiplication by 
$exp[i{\bf k}_*\cdot ({\bf x}-{\bf x}_*)-i\omega t]$) they connect smoothly 
onto the incoming and outgoing WKB wavefunctions.  

The uncoupled polarizations are used to reduce the full $N\times N$ 
dispersion matrix to the following $2\times 2$ {\em reduced dispersion matrix}
\begin{equation}
{\bf D}({\bf z})=
\left(
\begin{array}{cc}
D_{\alpha \alpha}({\bf z})   & D_{\alpha \beta}({\bf z})  \\
D_{\alpha \beta}^{*}({\bf z}) & D_{\beta \beta}({\bf z}) \\
\end{array}
\right)
\end{equation}
with $D_{jk}({\bf z})\equiv {\hat {\bf e}}^{\dag}_j\cdot {\bf D}({\bf z})\cdot {\hat {\bf e}}_k$ 
$j,k=(\alpha ,\beta )$.   We now consider only the reduced dispersion 
matrix (which is also Hermitian).  
Following Littlejohn and 
Flynn~\cite{robertandgregannals,robertandgregchaos,robertPRL}, 
we use the fact that any $2\times 2$ 
Hermitian matrix can be expanded using the Pauli matrices as a basis:
\begin{equation}
{\bf D}({\bf z})\equiv B_{\mu}({\bf z})\sigma ^{\mu}=
\left(
\begin{array}{cc}
B_0({\bf z})+B_3({\bf z}) & B_1({\bf z})+iB_2({\bf z}) \\
B_1({\bf z})-iB_2({\bf z}) & B_0({\bf z})-B_3({\bf z}) 
\end{array}
\right).
\end{equation}
Here, the components of the `four-vector' $B=(B_0,B_1,B_2,B_3)$ are {\em real} 
scalar functions of $\bf z$ which are assumed to be
independent in the region of interest.  Taking the determinant gives
$det({\bf D})=B_0^2-B_1^2-B_2^2-B_3^2 = \eta ^{\mu \nu}B_{\mu }B_{\nu}$,
with the Minkowski tensor $\eta \equiv diag (1,-1,-1,-1)$.  
Note that $det({\bf D})=0$ implies that $B$ must lie 
on the `light' cone in `$B$-space'.

Rays are propagated on the dispersion surface using 
$D({\bf z})=B^{\mu}B_{\mu}({\bf z})$ 
as the ray Hamiltonian. Hamilton's equations are most compactly written in 
terms of the Poisson bracket.  For 
any two scalar functions, $f({\bf z})$ and $g({\bf z})$, the Poisson bracket 
is defined to be $\{f,g\}\equiv \nabla _zf\cdot J\cdot \nabla _zg$ where
the $6\times 6$ matrix $J$ is defined as
\begin{equation}
J\equiv 
\left(
\begin{array}{cc}
0 & 1 \\
-1 & 0
\end{array}
\right)
\end{equation}
with `$0$' and `$1$' the $3\times 3$ null and identity matrices, respectively.
The ray Hamiltonian $D({\bf z})$ generates the ray evolution equations 
{\em via}:
\begin{equation}
{\dot {\bf z}}\equiv {{d{\bf z}}\over d\sigma }=\{D,{\bf z}\}
=-J\cdot \nabla _z D. 
\label{rayequationsz}
\end{equation}
Thus, ${\dot {\bf x}}=-\nabla _kD$ and ${\dot {\bf k}}=\nabla _xD$.
Following a ray, any scalar function
$f({\bf z})$ changes as 
${\dot f}=\{D,f\}=\{\eta ^{\nu \rho}B_{\nu}B_{\rho},f\}$.
In particular, the coordinates in `B-space' change as
\begin{equation}
{\dot B_{\mu}}=2\Omega ^{\nu}_{\mu}B_{\nu},\qquad \mu = 0,1,2,3.
\label{Bdot}
\end{equation}
Here $\Omega ^{\nu}_{\mu}=\eta ^{\nu \rho}\Omega _{\rho \mu}$ is 
a $4\times 4$ matrix composed of all pairwise Poisson brackets of the 
components $B_{\mu}({\bf z})$:  
$\Omega _{\rho \mu}({\bf z})\equiv \{B_{\rho},B_{\mu}\}=-\{B_{\mu},B_{\rho}\}$.
The matrix $\Omega $ plays a fundamental role in the theory; since
all of its entries are Poisson brackets, the entire matrix is invariant under
canonical transformations.  Infinitesimal Lorentz transformations 
are generated by anti-symmetric matrices (meaning that the $4\times 4$ matrix
${\Lambda }(\sigma )=1+\sigma \eta \Omega $ 
satisfies 
${\tilde \Lambda }(\sigma )\eta \Lambda (\sigma)=\eta +{\cal O}(\sigma ^2)$).  
Hence, we can associate a one-parameter family of
Lorentz transformations with the ray propagation.  The key idea is that the 
equations~(\ref{Bdot}) are simply Hamilton's equations written in a 
non-canonical coordinate system. (Recalling that phase space 
is six-dimensional, and there are only four $B_{\mu}$, we must 
supplement~(\ref{Bdot}) by two further evolution equations for another pair 
of coordinates, as will be shown.)  Note also that, when 
$B\rightarrow B'=A B$ for {\em any} constant linear transformation $A$
(not just a Lorentz transformation), the matrix of Poisson brackets 
$\Omega $ transforms via $\Omega '=A\Omega {\tilde A} $.  

The geometrical picture is as follows (see Figure (1)):  the 
ray phase space is six-dimensional and plays the role of the {\em base space}. 
Over each point $\bf z$ there is a fiber consisting of the 
space of $2\times 2$ Hermitian matrices (`$D$-space').  Equivalently, the 
fiber consists of the space of four-vectors $B$ (`$B$-space').  The fiber 
space is four-dimensional.  For each $\bf z$, assign a particular 
${\bf D}({\bf z})$ and assume that this assignment changes 
smoothly as we vary the base point ${\bf z}$.  This assignment defines a 
six-dimensional surface, denoted $\cal S$.   We note that a ray trajectory in 
the phase space (a solution of~(\ref{rayequationsz})) is associated with a 
well-defined curve in $\cal S$ and, hence, with a well-defined curve in the 
fiber space.  This curve obeys~(\ref{Bdot}) in $B$-space.

The mapping between the six-dimensional section $\cal S$ and the 
six-dimensional 
ray phase space is smooth and one-to-one.  However, the mapping from 
$\cal S$ to the four-dimensional $D$- and $B$-spaces cannot be one-to-one.  
This can be clarified by an appropriate choice of local coordinates. We have 
assumed that the four components $B_{\mu}({\bf z})$ are locally independent 
functions of 
${\bf z}$, hence we can use them as four local (non-canonical) coordinates in 
the ray phase space.  These are supplemented by two further 
independent coordinates, call them $F_1({\bf z})$ and $F_2({\bf z})$.  We
can choose $F_1$ and $F_2$ to satisfy
$\{F_1,B_{\mu}\}=\{F_2,B_{\mu}\}=0,\;\mu=0,\ldots 3$, but 
$F_1$ and $F_2$ are otherwise arbitrary (note that this implies 
$\{F_1,F_2\}\neq0$).  This gives us a local 
six-dimensional (non-canonical) coordinate system $(B_0({\bf z}),B_1({\bf z}),
B_2({\bf z}),B_3({\bf z}),F_1({\bf z}),F_2({\bf z}))$.  
The $B$-coordinates change {\em via}~(\ref{Bdot}), while the
$F_1$ and $F_2$ do not change along the ray since they have zero Poisson 
bracket with the $B_{\mu}$'s.  This implies that the rays lie in 
surfaces of constant ($F_1,F_2$) and, therefore, there is a natural foliation 
of the six-dimensional phase space, with each four-dimensional leaf 
labeled by the two invariants $F_1$ and $F_2$. Within each 
four-dimensional leaf, the three-dimensional image of the light 
cone $B^{\mu}B_{\mu}({\bf z})=0$ is a slice of the dispersion surface 
$D({\bf z})=0$.  Note that we do not need
to assume that we are near the apex of this cone, nor have we linearized
the coordinate functions.  These properties of the local ray dynamics
are purely a consequence of the fact that the ray Hamiltonian is the 
determinant of a generic $2\times 2$ Hermitian matrix.

We now fix attention upon a particular leaf ({\em i.e.} fixed values of 
($F_1,F_2$)).  After restricting to a fixed leaf, the ray motion can either 
be viewed in the phase space (where the ray equations~(\ref{rayequationsz}) 
are canonical), or in $B$-space (where the ray equations~(\ref{Bdot}) are 
non-canonical).  Extracting the geometrical invariants, and their physical 
implications, is more direct using the non-canonical coordinates.

Under a congruence transformation ${\bf Q}$ with constant entries, the 
$2\times 2$ reduced dispersion matrix transforms {\em via} 
${\bf D}\rightarrow {\bf D}'\equiv {\bf Q}^{\dag }\cdot {\bf D}\cdot {\bf Q}$,
and the determinant as $det({\bf D}')=Q^2det({\bf D})$ where 
$det({\bf Q})\equiv Q$.  First consider $\bf Q$ with unit determinant, 
$Q=1$.  Direct calculation shows that the related four-vector $B$ and 
matrix $\Omega $ transform to $B'_{\nu}=\Lambda ^{\mu}_{\nu}B_{\mu}$ and 
$\Omega '=\Lambda \Omega {\tilde \Lambda}$, respectively, where $\Lambda $ is
the $4\times 4$ Lorentz matrix with entries~\cite{robertandgregchaos,Weyl}
$(\Lambda ^{-1})^{\mu }_{\nu }=\frac{1}{2}tr\left( \sigma _{\mu}{\bf Q}^{\dag }\sigma _{\nu}{\bf Q}\right)$.  Notice that two ${\bf Q}$'s that differ by 
an overall minus sign are related to the same Lorentz transformation.

The Lorentz transformations leave the Minkowski tensor $\eta$
invariant ${\tilde \Lambda }\eta \Lambda =\eta $. The Minkowski tensor
satisfies $\eta ^2=1$, implying that 
$\Lambda ^{-1}=\eta {\tilde \Lambda}\eta $.  Therefore, 
$\eta \Omega $ transforms via the similarity transformation
$\eta \Omega '={\tilde \Lambda }^{-1}\left(\eta \Omega \right){\tilde \Lambda }$,
and its characteristic polynomial is invariant under Lorentz transformations. 
Some algebra shows that
$P(\lambda )=det(\eta \Omega -\lambda )=\lambda ^4 -\frac{1}{2}tr((\eta \Omega )^2)\lambda ^2 +det(\eta \Omega)$.
If we use the standard parametrization for generators of the Lorentz 
group~\cite{jackson}:
\begin{equation}
\Omega \equiv
\left( 
\begin{array}{cccc}
    0         &  \gamma _1  & \gamma _2    & \gamma _3  \\    
-\gamma _1    &      0      & -\omega _3   & \omega _2  \\
-\gamma _2    &  \omega _3  &     0        &-\omega _1 \\     
-\gamma _3    & -\omega _2  &  \omega _1   &      0 
\end{array}
\right)
\label{omega}
\end{equation}
then we find
$P(\lambda )=\lambda ^4 + (\omega ^2-\gamma ^2)\lambda ^2 -
({\gamma \cdot \omega})^2$.  Notice that $P(\lambda )$ depends only upon
$\lambda ^2$, hence the roots of $P(\lambda )=0$ come in plus-minus pairs.
Generically, $P(0)$ is strictly negative, while $P(\lambda ^2)$ is concave 
upward as a function of $\lambda ^2$.  Therefore, as a function of 
$\lambda ^2$, $P=0$ will have one negative root and one positive root.  This 
implies, in turn, that as a function of $\lambda $, $P=0$ will have a pair of 
$\pm$ pure imaginary roots, and a pair of $\pm$ real roots.  This implies that
the ray motion will be a combination of elliptic and hyperbolic behaviors.

This parametrization~(\ref{omega}) provides a useful physical 
interpretation for the
meaning of $\Omega $.  Fix a point ${\bf z}={\bf z}_0$ on the ray where
$\Omega _0\equiv \Omega ({\bf z}_0)$ is evaluated. 
Now use $\Omega _0$ to generate a one-parameter family of Lorentz 
transformations in $B$-space ($\sigma $ is the ray orbit parameter) 
$B(\sigma )=\Lambda (\sigma)B(0)\equiv exp(\sigma \eta \Omega _0)B(0)$. This 
provides a local approximation to the ray orbit in $B$-space.  In that space,
the ray evolution will be a combination of a rotation (generated by the 
$\omega $-subspace) and a `boost', generated by the 
$\gamma $-subspace~\cite{jackson}.  Hence, the orbit
in $B$-space will generically be a combination of elliptic and hyperbolic 
motions.  But, a ray orbit in $B$-space is the smooth image of a ray 
orbit in the original phase space, hence ray orbits generated by generic 
$2\times 2$ dispersion matrices will have a combination of elliptic and 
hyperbolic motions.  Purely hyperbolic motion is not possible.  

The above discussion demonstrates that ray dynamics in a multi-dimensional
conversion is a combination of elliptic and hyperbolic motions (and 
degenerate versions of these motions such as occurs, for example, when
$\omega =0$).  However, we have not yet exhausted all 
possible congruence transformations.
Now consider congruence transformations that are pure scale 
transformations ({\em i.e.} diagonal matrices with 
$|det({\bf Q})|\equiv Q\neq 1$).
From $det({\bf D}')=Q^2det({\bf D})$ we have 
$\eta ^{\mu \nu}B'_{\mu}B'_{\nu}=Q^2\eta^{\mu \nu}B_{\mu}B_{\nu}$.
Thus the light-cone is invariant, but the numerical value of the
determinant off the light-cone can change.  Using $B'=\Lambda B$ we find that
$\Lambda $ now preserves the Minkowski tensor only to an overall scale factor:
${\tilde \Lambda }\eta \Lambda =Q^2\eta$, implying $\Lambda $
is not a Lorentz transformation, but a {\em conformal} one.  
We still have $\Omega '=\Lambda \Omega {\tilde \Lambda }$, and therefore
$\eta \Omega '=Q^{2}{\tilde \Lambda}^{-1}\eta \Omega {\tilde \Lambda}$.
This implies that the characteristic polynomial of $\eta \Omega '$ is
$P'(\lambda )\equiv det(\eta \Omega '-\lambda )=
det({\tilde \Lambda}^{-1}(Q^{2}\eta \Omega -\lambda ){\tilde \Lambda})
=Q^8P(\lambda /Q^2)$.  Therefore
$P'(\lambda )=\lambda ^4 + (\omega ^2-\gamma ^2){\lambda ^2}{Q^4} -
{({\gamma \cdot \omega})^2}{Q^8}$,
and we arrive, finally, at the result that the ratio 
\begin{equation}
K \equiv {{\omega ^2-\gamma ^2}\over {\gamma \cdot \omega}}
\end{equation}
is the sole quantity that is invariant under all (constant) congruence 
transformations of the $2\times 2$ reduced dispersion matrix.
As mentioned earlier, the entries of 
$\Omega$ are Poisson brackets, hence the entire matrix is invariant under 
canonical transformations, and we have uncovered a new quantity that is 
invariant under both sets of transformations.

In a separate paper we shall discuss how to exploit the connection with the
Lorentz group to construct the {\em normal form} of the reduced dispersion
matrix.  If $\gamma $ and $\omega $ are not already parallel, it is possible
to make them so by performing a `boost' in the direction 
$\omega \times \gamma$ with an appropriate choice of boost parameter. Hence, 
there is a set of frames where 
$\omega $ and $\gamma $ are parallel.  In such a frame, the invariant we have 
uncovered reduces to a simple function $K=\kappa - \kappa ^{-1}$ 
with $\kappa \equiv \omega '/\gamma '$.
The primes indicate that these are measured in these special frames.
This reveals that the invariant $\kappa$ has a natural interpretation as the 
helicity of a ray and measures the rate at which it `rotates' relative to
the rate of exponentiation.  The sign of $\kappa $ (the `handedness' of the 
ray helix, as defined in the four-dimensional $B$-space) is also invariant.
We shall also discuss
how to solve the related $2\times 2$ wave equation using 
generalizations of the fourier transform, and how to incorporate these
new results into numerical ray tracing codes.

\acknowledgements{This work was supported by the US Dept. of Energy, Office 
of Fusion Energy Sciences.  We would like to thank Robert Littlejohn for 
helpful suggestions and comments.}

\pagebreak

\begin{figure}
\epsfxsize=3.3 truein
\centerline{\epsfbox{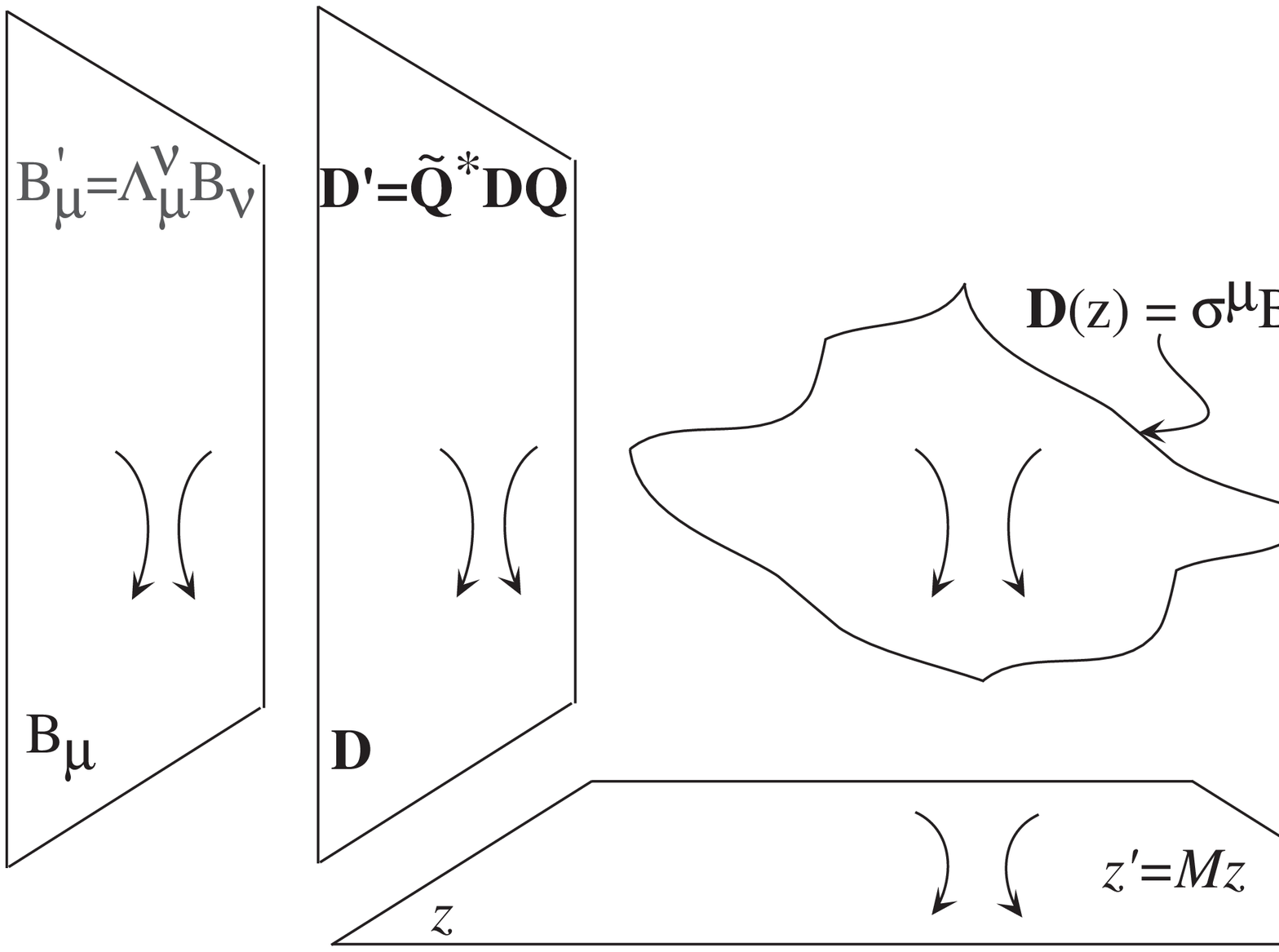}}
\vspace{.25in}
\caption{A conceptual figure showing the relationship between the 
six-dimensional ray phase (represented by the plane at the bottom of the 
figure), the four-dimensional fiber spaces (represented by the `$D$' and `$B$' 
planes at the left), and the six-dimensional surface 
${\bf D}({\bf z})\equiv {\cal S}$.  Note that a ray in the phase space defines 
a smooth curve in $\cal S$ and in the fiber spaces.}
\label{fig:fiberbundle.fig}
\end{figure}

\end{document}